\documentclass{osa-article}

\journal{osajournal}

\articletype{Research Article}

\usepackage{color}
\usepackage{ulem}

\makeatletter
\newcommand{\subscripts}[3]{%
  \@mathmeasure\z@\displaystyle{#2}%
  \global\setbox\@ne\vbox to\ht\z@{}\dp\@ne\dp\z@
  \setbox\tw@\box\@ne
  \@mathmeasure4\displaystyle{\copy\tw@_{#1}}%
  \@mathmeasure6\displaystyle{{#2}_{#3}}%
  \dimen@-\wd6 \advance\dimen@\wd4 \advance\dimen@\wd\z@
  \hbox to\dimen@{}\mathop{\kern-\dimen@\box4\box6}%
}
\makeatother

\newcommand{\tens}[1]{%
  \mathbin{\mathop{\otimes}\limits_{#1}}%
}

\begin{document}

\title{Memory-based Probabilistic Amplification of Coherent states}

\author{Keiichiro Furuya,\authormark{1} Mahdi Hosseini,\authormark{2,3*} }

\address{\authormark{1}Department of Physics and Astronomy, Purdue University, West Lafayette, IN 47907, USA \\
\authormark{2}Birck Nanotechnology Center, School of Electrical and Computer Engineering, Purdue University, West Lafayette, IN 47907, USA\\
\authormark{3} Purdue Quantum Science and Engineering Institute, Purdue University, West Lafayette, Indiana 47907, USA\\}


\email{\authormark{*}mh@purdue.edu} 



\begin{abstract}
	We analytically show that probabilistic amplification of a weak coherent state stored inside an atomic medium can be achieved via detection of coherently scattered photons. We show that this is because of collective excitations created among atoms in the ensemble. We describe the physics of the amplification and identify the failure events, which occur during the amplification process. The amplification is realized by coherently mapping a weak coherent state in an ensemble of $\Lambda$-level atoms followed by detection of multiple Raman scattered photons, conditionally projecting the coherent state into an amplified state upon retrieval. 
\end{abstract}

\section{\label{sec:level1}Introduction}

	Quantum state preparation and manipulation via continuous observation and projective measurement on a system of particles has led to realization of nontrivial quantum states \cite{Ourjoumtsev:2007aa, Bimbard:2010aa, McConnell:2015aa, Deleglise:2008aa, PhysRevLett.112.170501, Pfaff:2012aa}. The nonlinearity imposed by the measurement has been used to create entanglement between separate atomic ensembles \cite{Kimble4QM}, or many atoms within one ensemble\cite{ McConnell:2015aa, Pfaff:2012aa}. Another application of the projective measurement is in a noiseless amplification of quantum optical states \cite{ralund2008, pegg, xiang, zava, caves2013}. The probabilistic noiseless amplification has applications in state discrimination \cite{Barnett:09, PhysRevA.93.062315}, entanglement purification\cite{xiang} and quantum communication\cite{Chrzanowski:2014aa}. Probabilistic noiseless linear amplification (NLA) in optics, have been theoretically proposed \cite{ralund2008, pegg} and experimentally realized for weak coherent states of light \cite{xiang, zava}.

	In this paper, we propose a memory-based probabilistic amplification   as a new approach to amplification of  quantum information . We show that photons spontaneously scattered from an ensemble of atoms storing a coherent state of light, can project the collective atomic excitations to an amplified state. We show that the parameters of the probabilistic amplification including success probability and noise are tunable via the external fields.
	
	 In the case of optical states, applying $\hat{a} \hat{a}^{\dagger}$ operations to a weak coherent state, written as $|\alpha\rangle \approx |0>+\alpha |1>$, can approximately realize the amplification process, where $\hat{a}$ and $\hat{a}^{\dagger}$ are annihilation and creation operators, respectively. Experimentally, $ \hat{a}^{\dagger}$ process, can be realized by, for example, seeding a nonlinear crystal with the coherent state undergoing spontaneous parametric down conversion in the crystal\cite{zava}. Detection of an idler photon at the output of the crystal conditions the photon addition process. Subsequent subtraction of a single photon using a low-reflective beam splitter heralds the annihilation process. Mathematically, we can write 
	\begin{equation}   \label{eq:optamp}
		\begin{split} 
			\hat{a} \hat{a}^{\dagger} | \alpha \rangle \simeq &  \hat{a} \hat{a}^{\dagger}| 0 \rangle + \alpha  \hat{a} \hat{a}^{\dagger} | 1 \rangle \\
				= & | 0 \rangle +   2  \alpha  | 1 \rangle   \simeq  | 2\alpha \rangle.
		\end{split}
	\end{equation}
	
	In the current paper, we demonstrate the possibility of such amplification process in an atomic medium (quantum memory). The optical state forms a stationary excitation inside the memory while the nonlinearity needed for $\hat{a}^{\dagger}$ and $\hat{a}$ operation is provided by the  controlled Raman  scattering processes followed by photon detection. The collective excitations created upon detection of scattered single photons give rise to the amplification process. The collective excitation in atomic ensembles \cite{dicke1954} is the key concept in understating many phenomena in atomic physics including quantum storage\cite{Tanji:PRL2009}, bi-photon generation\cite{vuleticscience}, and entanglement creation\cite{Kimble4QM}. Such collective effects when combined with projective measurements have applications in quantum communication as proposed by Duan, Lukin, Cirac and Zoller (DLCZ) \cite{dlcz}.

			\begin{figure}[!t]
	 			 \begin{center}
			  		\includegraphics[clip,width=\columnwidth]{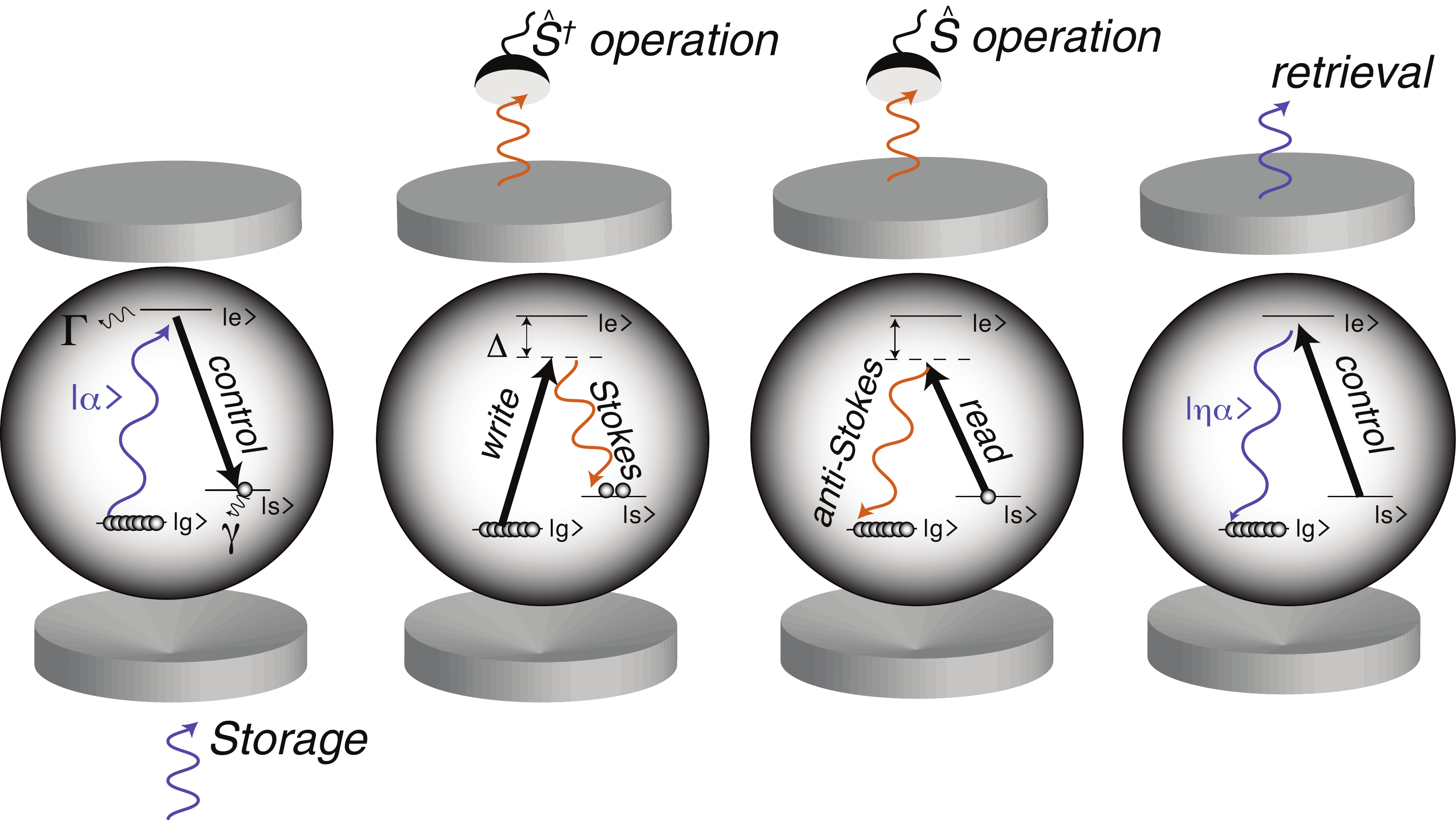}
   					\caption{Quantum State Amplification(Process) : $\hat{S}^{\dagger}$ and $\hat{S}$ processes are shown on the right, and left respectively. When pumping laser comes into the system, one of the atoms is excited to $|s\rangle $ via a spontaneous Raman process and emits a Stokes photon. This process corresponds to $\hat{a}^{\dagger}$ operation to the optical field. The $\hat{S}$ operation corresponds to a reversed process via which an anti-Stokes photon is created. This process resembles to $\hat{a}$ operation to the optical field.}
    					\label{fig:ampe}
  				\end{center}
			\end{figure}

	As quantum memories and NLA are important building blocks of future quantum communications, it is important to realize both processes in a compatible and efficient way. In our approach, both storage and amplification processes can be achieved in the same platform. The proposed amplification protocol is universal and can be applied to various quantum memories working based on $\Lambda$-level atoms\cite{Vernaz-Gris:2018aa, RamanQM, Hosseini:NatPhys:2011}. Deriving Raman transitions using a pair of ``write" and ``read" lasers and detecting photons entangled with the atomic ensemble gives rise to creation and amplification of coherent excitations. We show that such memory-based amplification scheme enables transduction and amplification of optical coherent states with applications in the future quantum communication networks\cite{Gisin:Nature2007, Kimble:Nature2008, Liao:2017aa}.

	The remainder of the paper proceeds as follows. In section II, we provide a general discussion of the amplification of weak coherent states. In section III, we discuss the basic processes involved in our scheme. In section IV, we introduce the concept of quality of the amplification and identify the sources of the failure of the process. Section V briefly discusses the implementations of the amplification. In section VI, concludes the discussions.

\section{\label{sec:awcs}Storage and  Amplification of Optical States}

Faithful  storage of quantum optical states in atomic media has been achieved relying on different storage mechanisms including electromagnetically induced transparency(EIT)~\cite{Fleischhauer-EIT, Vernaz-Gris:2018aa}, off-resonance Raman \cite{Reim:2010p11902}, and controlled reversed inhomogeneous broadening \cite{Afzelius:2009p11906, Hedges:2010p11910, Hosseini:NatPhys:2011}. When a weak coherent optical state, i.e. $ | \alpha \rangle = | 0 \rangle + \alpha |1 \rangle $, is stored as a collective atomic excitation, the optical coherence is mapped to the coherence between two atomic states. In the case of $\Lambda$-level atom shown in Fig.\ref{fig:ampe}, the collective atomic excitation after storage can be written as $| G \rangle_A + \alpha | S \rangle_A$, where
	
	\begin{equation}   \label{eq:2}
			\begin{split}
				|G \rangle_A = & | g_1 g_2 \cdots g_{N}  \rangle \\
				|S \rangle_A = &  \frac{1}{\sqrt{N}} \sum_{i=1}^{N} | S_i  \rangle \\
					= & \frac{1}{\sqrt{N}} \Big\{ | s_1 g_2  \cdots  \rangle +  \cdots +  | g_1 \cdots s_{N}  \rangle  \Big\} ,  \\
			\end{split}
		\end{equation}
where $g_i$ and $s_i$ $(i=1,2, ..., N)$ belong to $i$th atom.  The state $| S_i \rangle  $ corresponds to a state where only $i$th atom is excited and the others are in the ground state.   Here, the subscript "A" denotes the atomic subspace. For example, $|G \rangle_A$ describes the ground state of the atomic ensemble. This notation will be used throughout the paper.   The stored information takes the form of a coherent superposition of atomic ground states and excited atomic state with relative probability amplitude $\alpha$. The mapping of the atomic excitation to the optical excitations is ensured by the reversibility of the storage protocols. For the rest of this paper, we assume that a reversible mapping from optical to atomic states is achievable and we only focus on describing the amplification process of the atomic excitations as an analogue to optical amplification. 

	In our case, the amplified atomic state (unnormalized) is $| G \rangle_A + \eta \alpha |  S \rangle_A$, where $\eta$ is the gain of the amplification process. This is analogous to the optical state amplification in Eq.\ref{eq:optamp}. The atomic state amplification is heralded by detection of two Raman scattered photons, Stokes and anti-Stokes photons. The Stokes and anti-Stokes photons are generated through a spontaneous Raman process as shown in Fig.~\ref{fig:ampe}. Based on these processes, a protocol for entanglement distribution has been proposed by DLCZ\cite{dlcz} and successfully implemented for four local ensembles of atoms \cite{Kimble4QM}. Moreover, quantum memory-based Raman scattered photon sources have been successfully demonstrated for 100s of modes \cite{Pu:2017aa, Parniak:2017aa} paving the way for the future quantum networks.

\section{\label{sec:level2}  Probabilistic Amplification via Atomic Coherence }

%
%
%
\subsection{ Basic Processes }
		
		We consider an ensemble of $N$ three-level atoms ($\Lambda$ system) as shown in Fig.\ref{fig:ampe}. The state $|g\rangle$ represents the ground state, and $|s\rangle$ is the meta-stable state of a single atom. The state $|e\rangle$ is the excited state via which atoms are transferred between $|g\rangle$ and  $|s\rangle$ states. As an example, the hyperfine ground levels of the alkali atoms interacting with one of the excited states via a two-photon transition (Fig.1) can be considered as the $\Lambda$ system.  For the collective atomic states, as in Eq.\ref{eq:2}, $| G \rangle_A $  represents all atoms in the ground state. Similarly, $| S \rangle_A $ represents only the $i$th atom is in the $|s\rangle$ state, otherwise in the $|g\rangle$ state. In general, the collective atomic state with $k$ atoms excited are described by Dicke state \cite{group}:

		\begin{equation} \label{eq:dicke}
			\begin{split}
				|k,N\rangle_A =  \sqrt{\frac{k!(N - k)!}{N !}}  \sum_{p}   | s_1 ... s_k g_{k+1} ... g_{N}  \rangle
			\end{split}
		\end{equation}
		where $\sum_{p}$ denotes sum of all permutations. As an example, Eq.\ref{eq:dicke} equals state $| G \rangle_A$ and  $| S \rangle_A$ when $k = 0$ and $k=1$, respectively.
		
		 To define transition operators between the two atomic energy levels, we adopt second-quantized atomic Hamiltonian formalism presented in Ref.\cite{loudon} as 	
		\begin{equation}   \label{eq:4}
			\begin{split}
				\hat{S} & \equiv  \frac{1}{\sqrt{N}} \sum_{i=1}^{N} | g \rangle_{i} \langle s |. \\     
			\end{split}
		\end{equation}
		  The operator $\hat{S}$, is the atomic lowering operator resulting in $i$th atom de-excitation ($|s\rangle\to|g\rangle$) creating a Stokes photon (in mode $a$) as the result. The conjugate of the atomic operator is $\hat{S}^{\dagger}$, which is responsible for raising the $i$th atom from $|g\rangle$ to $|s\rangle$ state, creating an anti-Stokes photons (in mode $b$).	 

		Fig.\ref{fig:ampt} explains how the coherent superposition of different atomic excitations plays a role in achieving gain, $\eta>1$. Considering the case of $k=1$, state $|S\rangle_A = | 1, N\rangle_A$ is a superposition of $|  S_i \rangle$ (single excitations) as in Eq.\ref{eq:2}. When  the operator $\hat{S}^{\dagger}$ acts on the state $|  S \rangle_A$, the probability corresponding to eigenstate, $|S_{i} S_{j} \rangle$, increases by a factor of two. This is because  both $|  S_i \rangle$ and  $|  S_j \rangle$ in the initial state are excited to the same final state, $|  S_{i} S_{j} \rangle$. We note that states $|  S_{i} S_{j} \rangle$ and $|  S_{j} S_{i} \rangle$ are indistinguishable. The indistinguishability of the path to the final state  is the key to the  amplification process. The final annihilation process, $\hat{S}$, reduces the two excitations, for instance $|  S_{i} S_{l} \rangle$ $( l = 2, \cdots , N)$, into the state with one excitation, $|  S_{i}  \rangle$.

		  The generation of two Raman photons (Stokes and anti-Stokes photons) in modes $a$ and $b$ are the key processes corresponding to $\hat{S}$ and $\hat{S}^{\dagger}$ operators, respectively. The optical states are described in the number basis where $ |0 \rangle_{a,b}$ is the optical vacuum number state of the two modes and operators $\hat{a}$ and $\hat{b}$ are the corresponding lowering optical operators. For simplicity, we drop the operator symbol, hat, from now on.  
		
 The photons are spontaneously generated by weak pumping lasers driving $|G\rangle_A \to |S\rangle_A$ (write pump) and $ |S\rangle_A   \to  |G\rangle_A $ (read pump) transitions.  The two read and write pumping processes are controlled by time, power and detuning of the lasers from the excited state. Detection of a scattered Stokes photon in the mode   $a$ whose spatio-temporal profile matches that   of the initial coherent state, registers creation of a collective excitation, which is mode-matched to the stored atomic coherence. Subsequently, driving the system with a read pump light, can result in emission of an anti-Stokes photon in mode   $b$ whose spatio-temporal profile also matches that of the initial coherent state    completing the amplification process. We note here that the retrieval of the anti-Stokes photon can not be deterministically achieved via the EIT excitation and needs to be also probabilistic to avoid retrieval of the coherent state. The Stokes and anti-Stokes processes of this kind have been proposed to create quantum repeater networks \cite{dlcz} and also efficient generation of photon pairs in this way has been realized using both room temperature atoms \cite{eisaman-eit-sph} and cold atoms in optical cavities\cite{vuleticscience}. The emitted Stokes photon is entangled with the collective excitations of the atoms reversibly mapped to an anti-Stokes photon.  

  \subsection{ Interaction Hamiltonians and state evolution}  
	The interaction Hamiltonians and their corresponding creation and annihilation operators giving rise to the Stokes and anti-Stokes  photon creation, $S^{\dagger}$ and $S$ processes, are generally
given by		
		\begin{equation} \label{eq:hw1}
			H_W(t) = i \sqrt{p_w}  S^{\dagger} a^{\dagger}(t) + h.c. 
		\end{equation}
		\begin{equation} \label{eq:hr1}
			H_R(t) = i \sqrt{p_r} S b^{\dagger}(t) + h.c. 
		\end{equation}
  The Stokes and anti-Stokes process gives rise to creation of photons in modes $a$ and $b$ whose corresponding creation operators are given by \cite{3d}: 
		 		\begin{equation} \label{eq:a}
			a^{\dagger} (t) =  - i  \frac{\sqrt{N}}{\sqrt{p_{w}}}   \int g_{w} a_{\mathbf{k}}^{\dagger} e^{i\Delta \omega_{a} t }  \langle u_w(\mathbf{r}_{i}, t) e^{-i\mathbf{\Delta k_w} \cdot \mathbf{r}_{i}}\rangle d^{3} \mathbf{k} 
		\end{equation}
	
		\begin{equation}  \label{eq:a_As}
			b^{\dagger} (t) = - i  \frac{\sqrt{N}}{\sqrt{p_{r}}}  \int g_{r} b_{\mathbf{k}}^{\dagger} e^{i\Delta \omega_{b} t }  \langle u_{r}(\mathbf{r}_{i}, t) e^{-i\mathbf{\Delta k_r} \cdot \mathbf{r}_{i}}\rangle d^{3} \mathbf{k} 
		\end{equation}
		 where $\mathbf{r}_{i}$ is the coordinate of the i-th atom, $g_{w}$, $g_{r}$ are light-atom coupling rates that are determined by transition strength involved in the writing and reading processes, the frequency terms $\Delta \omega_{a} = \omega_{\mathbf{k}} - \omega_{w} + \omega_{sg}$ and $\Delta \omega_{b} = \omega_{\mathbf{k}} - \omega_{r} - \omega_{sg}$, where $\omega_{sg} = \omega_{s} - \omega_{g}$ is the frequency difference between the ground and metastable excited atomic states. The angular frequencies $\omega_{w}$ and $\omega_{r}$ are the frequency of write and read pumping lasers, respectively. Moreover, $\mathbf{\Delta k_w} = \mathbf{k} - k_w\mathbf{z}$ and $\mathbf{\Delta k_r} = \mathbf{k} - k_r\mathbf{z}$, where $\mathbf{z}$ is the unit vector in the propagation, $z$, direction. The envelope functions $u_w(\mathbf{r}_{i}, t)$ and $u_{r}(\mathbf{r}_{i}, t)$ are the amplitude of pumping lasers for both writing and reading processes, respectively. The notation  $ \langle \cdots \rangle $   indicates taking an average over all the position variables $r_{i} $ . This averaging is valid for non-interacting atoms when atoms in $r_{i}$ and $r_{j}$ positions are uncorrelated and the probability distribution for each atom is the same and is determined by the geometry of the atomic medium. The coupling terms $p_i$ $(i = w,r)$ can be derived from the transition strength and intensity of the pumping lasers as:
		\begin{equation} \label{eq:pw}
			\begin{split}
				p_{w} = & N \int_{0}^{t_{0}} \int_{0}^{t_{0}} d\tau_{1} d\tau_{2} \int_{0}^{\mathbf{k_D}} d^3 \mathbf{k}  | g_{\mathbf{k}} |^2  e^{i\Delta \omega_{a}  ( \tau_1  -  \tau_2 )  } \\
					& \times | \langle u_a(\mathbf{r}_{i}, \tau_1 ) e^{-i \Delta \mathbf{k_w} \cdot \mathbf{r}_{i}}  \rangle |^2    \\
			\end{split}
		\end{equation}
		
		\begin{equation} \label{eq:pw}
			\begin{split}
				p_{r} = & N \int_{0}^{t_{0}} \int_{0}^{t_{0}} d\tau_{1} d\tau_{2} \int_{0}^{\mathbf{k_D}} d^3 \mathbf{k}  | g_{\mathbf{k}} |^2  e^{i\Delta \omega_{b}  ( \tau_1  -  \tau_2 )  } \\
					& \times | \langle u_{b}(\mathbf{r}_{i}, \tau_1 ) e^{-i \Delta \mathbf{k_r} \cdot \mathbf{r}_{i}}  \rangle |^2    \\
			\end{split}
		\end{equation}

		where $\tau_i (i = 1, 2)$ is the pulse duration and $\mathbf{k_D}$ determines the range of modes detected. 
			
	To find the unitary operators corresponding to the writing and reading Hamiltonians we define $U_W = e^{i \frac{H_W}{\hbar} t}$ and $U_R= e^{i \frac{H_R}{\hbar} t}$, assuming that $p_i$ $(i = w,r)\ll1$ and thus terms with higher order photon numbers are negligible, we can find the first order final state by applying time-dependent perturbation theory. Based on the first order perturbation, the unitary operators are
	\begin{equation} \label{eq:hw1}
		U_W = 1 - i \int H_W(\tau) d\tau =  (1 +  \sqrt{p_{w}} S^{\dagger} \tilde{a}^{\dagger} -  \sqrt{p_{w}} S \tilde{a}) 
	\end{equation}
	\begin{equation} \label{eq:hr1}
		U_R = 1 - i \int H_R(\tau) d\tau =  (1 +  \sqrt{p_{r}} S \tilde{b}^{\dagger} -  \sqrt{p_{r}} S^{\dagger}  \tilde{b}) 
	\end{equation}
The time evolution of the initial state, $|k,N\rangle_A \tens{} | 0 \rangle_a | 0 \rangle_{b} $, where $k$ number of atoms out of $N$ atoms are initially excited is described as :
	\begin{equation}    \label{eq:1stord}
		\begin{split} 
			 U_R U_W & |k,N\rangle_A \tens{} | 0 \rangle_a | 0 \rangle_{b} \\
			 = &    (1 +  \sqrt{p_{r}} S \tilde{b}^{\dagger} -  \sqrt{p_{r}} S^{\dagger}  \tilde{b})  \\
			 \times&  (1 +  \sqrt{p_{w}} S^{\dagger} \tilde{a}^{\dagger} -  \sqrt{p_{w}} S \tilde{a}) |k,N\rangle_A \tens{} | 0 \rangle_a | 0 \rangle_{b}  \\
			 =& (1 + \sqrt{p_{w}} S^{\dagger} \tilde{a}^{\dagger} + \sqrt{p_{r}} S \tilde{b}^{\dagger} + \sqrt{p_{r} p_{w}} S S^{\dagger} \tilde{b}^{\dagger} \tilde{a}^{\dagger}  )  |k,N\rangle_A \tens{} | 0 \rangle_a | 0 \rangle_{b} 
		\end{split}
	\end{equation}
	{After detecting Stokes and anti-Stokes photons, the atomic state is projected to the last term in Eq.\ref{eq:1stord}.
			\begin{equation}  \label{eq:1stord2}  
			\begin{split} 
				& \big\{ {}_{b}\langle 0  | \tilde{b} \tens{}  {}_{a}\langle 0 | \tilde{a}  \big\}  U_R U_W |k,N\rangle_A \tens{} | 0 \rangle_a | 0 \rangle_{b}  \\ 
					& \propto \sqrt{p_{w} p_r}  S S^{\dagger}  |k,N\rangle_A  \\
					&=\sqrt{p_w p_r} (k+1)(1-\frac{k}{N}) |k, N \rangle_A.
			\end{split}
		\end{equation}
	We use $\propto$ because we did not consider the normalization. The time-averaged operator $\tilde{a}^{\dagger} = - i \int a_j^{\dagger}(\tau) d\tau$, and similarly defined for $\tilde{b}$.

		 Upon detection of the two Raman photons, the final state is projected to the state with the modified amplitude. This modified amplitude becomes the gain of the process, $\eta = (k+1)(1- \frac{k}{N})$, in the amplification of the atomic coherent state. Note that to ensure $\eta >1$, we need $N \geq k+2$, e.g. at least three atoms are required. Using Eq.\ref{eq:1stord2} one can define the atomic amplification operator as $S S^{\dagger}$ similar to the optical amplification operator, $a a^{\dagger}$ in Eq.\ref{eq:optamp}.

		\begin{figure}[!t]
 			 \begin{center}
			  	\includegraphics[clip,width=7.0cm]{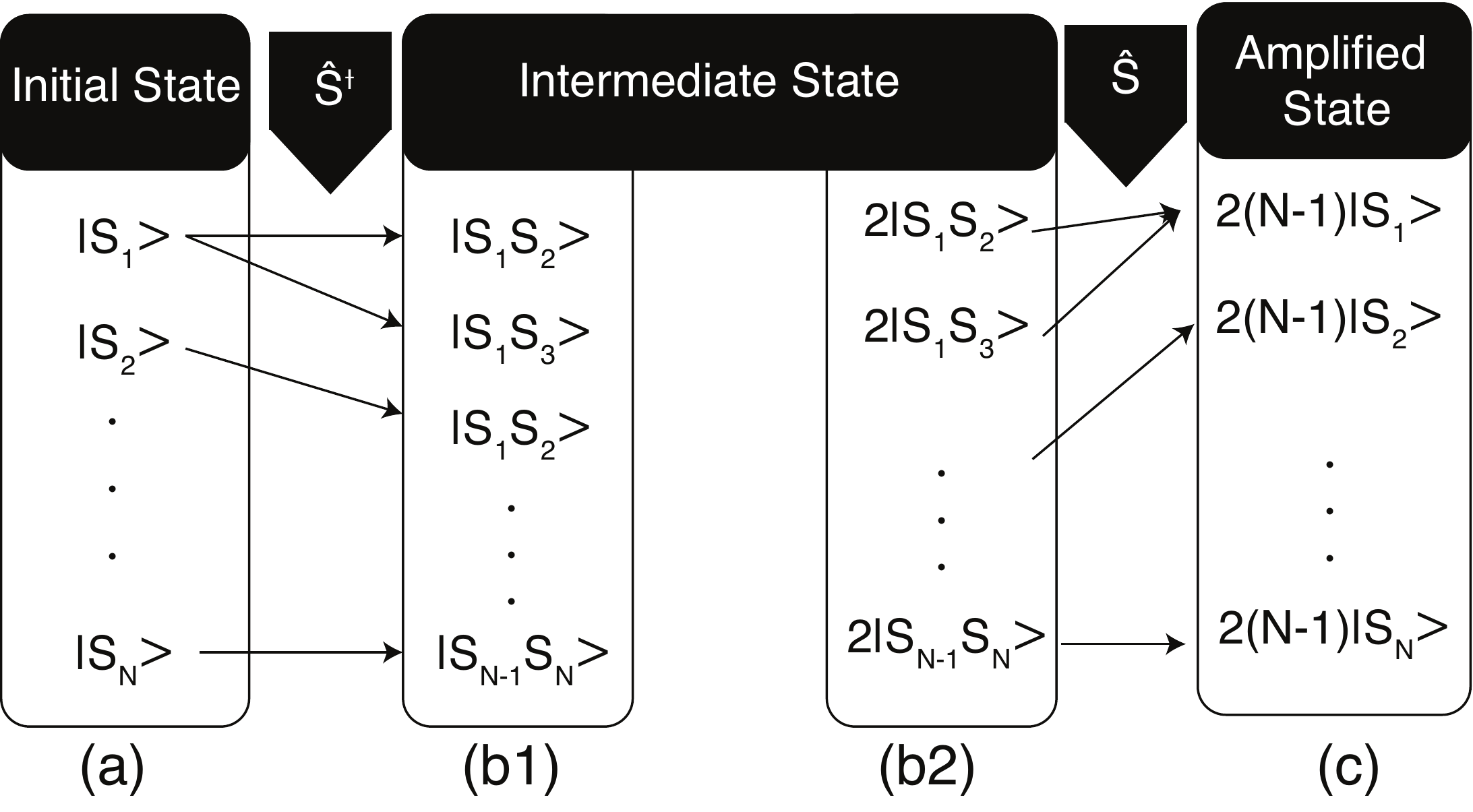}
   				\caption{The principle of quantum amplification: (a) The creation operator $\hat{S}^{\dagger}$ operates on the initial collective excitation described by $|S_j \rangle$ in the first column. (b1)(b2) Some of the resulted states are degenerate accounting for the factor 2 in the third column. (c) The final annihilation process, $\hat{S}$, reduces the two excitations into one with a modified probability (unnormalized). The coefficient $N-1$ appears because there are $N-1$ degenerate states after the $\hat{S}$ operation. }
    				\label{fig:ampt}
  			\end{center}
		\end{figure}

	\subsection{Multiple amplification processes}

		In this section we describe the amplification gain for multiple amplification processes. We note that  $S $ and $S^{\dagger} $ do not commute. Thus, we have two types of amplification operators to achieve the multiple amplification processes. The first process, type-I, is to repeat the amplification process $n$ times, i.e. $(S S^{\dagger} )^n$. The second process, type-II, is to repeat $ (S^{\dagger})^n$ first, and $ S^n$ second. Then, the gain of each type is 
		\begin{equation} \label{eq:type1}
			( S S^{\dagger} )^n  |k,N\rangle_A   = (k+1)^n (1- \frac{k}{N})^n |k,N\rangle_A,
		\end{equation}
		
		\begin{equation} \label{eq:type2}
			( S)^n  (S^{\dagger} )^n  |k,N\rangle_A   =  \prod_{h = k + 1}^{k + n} h ( 1  - \frac{h-1}{N}  ) |k,N\rangle_A .
		\end{equation}

				  In particular, let us consider an atomic state in a coherent superposition of the $|G\rangle$ and $|S\rangle$ states with full initial state written as  $|\Psi _{in}\rangle =  |\Psi _{in}\rangle_A \tens{} |0 \rangle_{b} |0 \rangle_{a} $ where $|\Psi _{in}\rangle_A =| G \rangle_A    +\alpha | S \rangle_A$ is the initial (unnormalized) atomic state. The coefficient   $\alpha$ is the amplitude of the coherent atomic state. The atomic state in this form can be created after an optical coherent state with amplitude $\alpha$ is mapped to a collective atomic state initially prepared in $|G\rangle_A$ as mentioned earlier. Then, the amplified state becomes
		 \begin{equation}    \label{eq:2stord}
			\begin{split} 
				 &{}_{b}\langle 0  | b \tens{}  {}_{a}\langle 0 | a U_R U_W |\Psi_{in}  \rangle  \\
					& \propto \sqrt{p_{w} p_r}  (   | G   \rangle_A   +  2\alpha  (1 -  \frac{1}{N})  | S  \rangle_A  ) .  \\
			\end{split}
		\end{equation}

		The successive measurements on the two Raman photons project the final state to an amplified atomic coherence   with a gain of nearly two. In this way, we can describe the states going through different amplification processes. For type-I amplification, we can write the final pure atomic state as 
		\begin{equation}    \label{eq:6}
			\begin{split}
				| \Psi_{amp} \rangle_{I} =& (S S^{\dagger})^n | \Psi_{in} \rangle_A  =  \Theta_I \Big\{ | G \rangle_A +  2^n ( 1 - \frac{1}{N}  )^n \alpha |  S \rangle_A  \Big\} ,\\     	     
			\end{split}
		\end{equation}
		where  $\Theta_I$ is a normalization constant. The gain in this case is $2^n ( 1 - \frac{1}{N}  )^n $.
		For the type-II amplification, ideally the final atomic state has the following form 
		\begin{equation}    \label{eq:7}
			\begin{split}
				| \Psi_{amp} \rangle_{II} =& S^n (S^{\dagger})^n | \Psi_{in} \rangle_A    =   \Theta_{II}  \Big\{ | G \rangle_A +  (n+1) ( 1  - \frac{n}{N}  )   \alpha |  S \rangle_A  \Big\} ,\\    
			\end{split}
		\end{equation}
		where  $\Theta_{II}$ is a normalization constant. The gain for this case is $(n+1)(1-\frac{n}{N})$. The gain of type-I amplification grows exponentially with $n$ compared to the linear growth in type-II (see Fig.\ref{fig:multamp2}). When considering multiple amplification trials, one needs to consider that noise is also accumulated during multiple stages of the amplification degrading the state fidelity. The detection of every photon pair marks the completion of one amplification stage however the produced atomic state as the result of such detection might not correspond to the desired amplified state. Therefore, the gain should not be considered the sole indicator of the amplification process and failure events degrading the state fidelity also needs to be considered when evaluating the performance of the amplifier. 		
		
%

		\begin{figure}[!t]
 			 \begin{center}
			  	\includegraphics[clip,width=7.0cm]{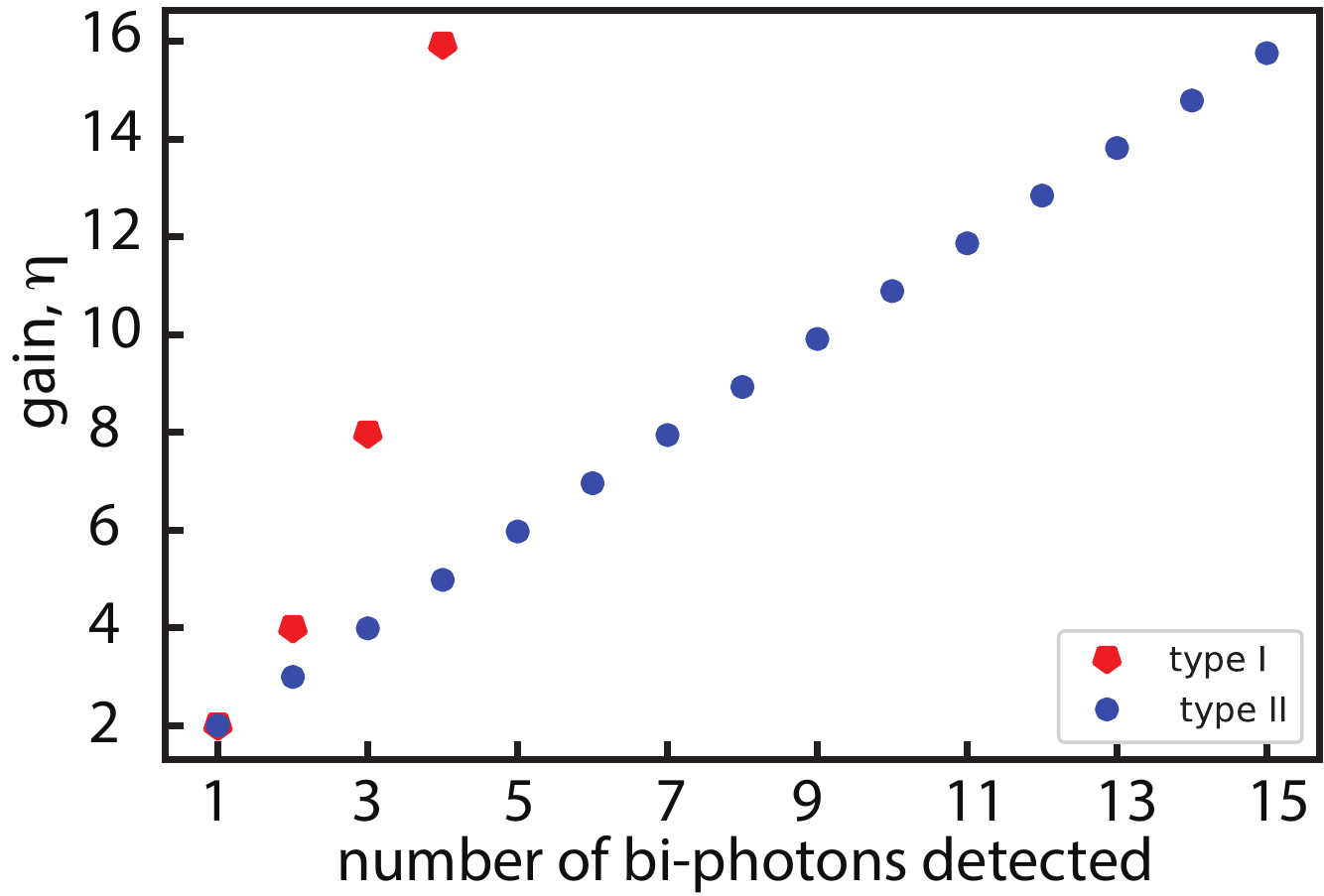}
   				\caption{ Plot of gain from Eqs. 18 \& 19 for multiple amplification processes showing the gain after applying two kinds of amplification operators, $(S S^{\dagger} )^n$ (type-I) and $S^n (S^{\dagger})^n$ (type-II). The horizontal axis indicates the number of amplification processes applied to the state or equivalently number of bi-photons detected during the amplification process.  }
    				\label{fig:multamp2}
  			\end{center}
		\end{figure}

\section{\label{sec:level3} Quality of Memory-based Probabilistic Amplification Process}
		
	In order to quantify the amplification process by taking into the account the failure events, we introduce the quantity and the quality of probabilistic amplification, $Q_{amp}$. This has been introduced by Pandey $et \ al.$ \cite{caves2013} for the characterization of probabilistic amplifiers, which is the product of success probability, $P_{suc}$, and fidelity, $F$, i.e. $Q_{amp}=P_{suc} \times F$. To express this quantity for the proposed amplification process, we derive the success probability, $P_{suc}$, based on the results of the previous section. Then, we clarify the sources of failure of our protocol, which degrade $Q_{amp}$.

		Consider the target state expressed in Eq.\ref{eq:1stord}. The success probability of the amplification process can then be calculated by taking into the account the other outcomes of the processes including the terms with only single and no photon detections. The success probability is then given by:
			\begin{equation} \label{eq:psuc1}
				P_{suc} = \frac{p_w p_r}{1 + p_w + p_r + p_w p_r + O(p^3)} .
			\end{equation} 	
			For this derivation, we consider up to the second order of $p_i$ $(i = w,r)$ including the cross terms, e.g. $p_w p_r $, in the denominator.   The probabilistic nature of the amplification has its origin in spontaneous Raman processes whose overall probability is determined by the light-atom coupling $p_w p_r$.   
			
	   The thermal nature of photon statistics when spontaneous Raman scattering is considered limits the overall success probability. The square of the coupling coefficients, $p_{r}$ and $p_w$, are directly related to the probability of generating Stokes and anti-Stokes photons in the detection mode. We note that in this calculation the detection efficiency is assumed to be unity and if not $p_{r/w}$ needs to be rescaled by the detection inefficiencies. Although in principle, one can consider higher-order terms when $p_{r/w}\not\ll1$ when they are not negligible, the resulted mixed-state after amplification will only make the process less desirable. Therefore, in free-space and when the scattering process is driven by a thermal bath, the  $p_{r/w}$ should be kept well below one such that higher order terms are negligible. We note that by engineering the environment of the atoms, directional emission can be achieved \cite{bohnet2012steady,goban2015superradiance, scheibner2007superradiance} significantly lowering the noise due to higher order terms. In such nonlinear scattering regime, the signal-to-noise ratio can be enhanced where less undetected photons are emitted. Study of the proposed amplification in the strong-coupling regime is interesting, however, it is beyond the scope of the current study. Therefore in this work, we focus on linear light-atom interactions. When considering  experimental realizations in this linear regime, the total probability of photons being emitted into all directions needs to be approximately $<0.1$ to ensure negligible higher order photon generation processes. The exact determination of value of the success probability is then found by the detection cone/mode which in part is determined by optimizing the failure processes.  
	
		The sources of failure of the process in the current system are higher-order processes or multi-photon processes, mode-mismatching loss, and spontaneous emission loss\cite{3d}. The higher-order processes are mitigated by reducing the pump power. The other two sources are more fundamental which depends on atomic geometry and optical mode. These two loss/noise processes are defined  similar to Ref.\cite{3d} and are attributed to mode-mismatching and spontaneous emission loss as
		
		\begin{equation}  \label{eq:pmode}
			P_{mode} = 1 - \frac{tr ( \langle \Psi_{amp} |   \rho_f    | \Psi_{amp}    \rangle  ) }{tr (  {}_{b}\langle 0 |{}_{a}\langle 0 |  b a  \rho_f  b^{\dagger} a^{\dagger} |  0  \rangle_{b} |  0  \rangle_a  )}
		\end{equation}
		
		\begin{equation}
			P_{spon} = 1 - \frac{tr ( \langle \Psi_{amp} |   \rho_f    | \Psi_{amp}    \rangle  ) }{tr (  {}_A\langle \Psi_{amp} |  \rho_f | \Psi_{amp}    \rangle_A  )}
		\end{equation}
		where $\rho_f$ is a final density matrix after the amplification.
		The states $| \Psi_{amp}    \rangle$ and $ | \Psi_{amp}    \rangle_A$  correspond to the total and atomic part of the amplified states, respectively.

		The spontaneous emission loss represents the probability of not detecting photons in mode $b$ and $a$ even when an atomic mode is created by the amplification operator, $S S^{\dagger}$. Similarly, the mode-mismatching loss represents the probability of not creating an atomic excitation upon applying $S S^{\dagger}$ even when photons are detected in the mode $b$ and $a$. The spontaneous emission loss is directly related to the success probability. On the other hand, the inherent mode-mismatching loss in the system limits the fidelity. These undesired effects impose a fundamental lower limit to the performance  of the system, like any other process relying on spontaneous processes.   The success probability of NLA process using down conversion in a nonlinear crystal is on the order of $10^{-5}$, taking into the account a pair production rate of kHz, pulse duration of $\mu$s and a 99/1\% beamsplitter (used for photon subtraction). In these schemes, although the fidelity can be close to unity the low success probability does not provide quantum advantage \cite{caves2013}. We note that measurement-based approaches can also be used for NLA to extract quantum information using post processing approaches \cite{Chrzanowski:2014aa}.  In our case, the  noise is introduced in expense of increasing the success probability. 	Because the probability and noise processes depend on the geometry of atomic cloud and detection modes, one needs to optimize both to achieve enhanced performance. The major noise or loss processes of spontaneous photon generation from an atomic cloud is discussed in details in Ref.\cite{3d} and proper geometry and configuration were identified for optimum free-space interactions. We note that the proposed protocol may introduce more noise when compared with the previously demonstrated NLA schemes \cite{xiang, zava}, however, because the amplification is performed on stationary excitations the effective success probability can be enhanced. As both success probability and fidelity are crucial \cite{caves2013} in evaluating the amplification process we introduce a metric, amplification quality, which is the product of these quantities.

		We can rewrite the success probability as $P_{suc}=tr ( \langle \Psi_{amp} |   \rho_f    | \Psi_{amp}    \rangle  )$, and moreover it can be shown that $P_{suc}$ is proportional to $1-P_{spon}$. The fidelity of the amplification process is directly related to the $P_{mode}$ and can be written as $1 - P_{mode}$. Thus the quality, $Q_{amp}$, is defined as 
		\begin{equation} \label{eq:qamp}
			\begin{split}
				Q_{amp}= & P_{amp} \times (1- P_{spon}) (1 - P_{mode}) \\
			\end{split}
		\end{equation}
		where $P_{amp} = tr (  {}_A\langle \Psi_{amp} |  \rho_f | \Psi_{amp}    \rangle_A  )$ is the probability of finding the atomic states to be in the desired amplified state,  $| \Psi_{amp}    \rangle_A =|G\rangle_A +2\alpha |S\rangle_A$, when $N\gg1$.   This probability is close to one where only the first-order photon generation processes are considered. However, as the amplification process is influenced by the higher-order processes at the elevated pump powers, the probability $P_{amp}$ decreases below one. Furthermore, the quality of amplification is determined by the intrinsic mismatch between atoms and detection modes as well as scattering into the free space. Therefore, the optimum performance is achieved when $P_{mode} = P_{spon}$ is minimized \cite{3d}.

\section{\label{sec:level4} Potential realization}
	The proposed memory-based amplification can be realized in an ensemble of $\Lambda$-level atoms coherently interacting with laser light. In free space, the loss terms $P_{mode}$ and $P_{spon}$ are limited by the free-space scattering projecting the atomic state into an undesired mode. In this regime, as shown in Ref. \cite{3d}, an optimum value for the loss processes can be achieved by a careful choice of  the detection modes. 
	As shown in Ref.\cite{3d},   the optimum loss probabilities for a single process (Stokes process) is about 30\% for spontaneous emission as well as the inherent mode-mismatching noise probabilities. It was suggested that the optimum limit for an atomic gas ensemble can be achieved by selective detection using an aperture. Using multiplexed approaches in cold atoms, simultaneous detection of hundreds of correlated photons in multiple spatial modes has been observed \cite{Parniak:2017aa}. Such techniques can be leveraged to enhance the success probability while the loss processes are not affected.  In that case, to ensure mode-matching between the emitted photons (detection modes) and the initial coherent state (stored excitation), the input coherent state, to be amplified, also needs to be multiplexed into the multiple detection channels, in order to increase the amplification probability by the number of multiplexed channels. To decrease the scattering and mode-matching losses in a single-mode configuration, the state-of-the-art platforms with cold atoms trapped near   \cite{sorensen2016coherent} or inside\cite{Langbecker:2018aa, goban2015superradiance}  the fibers or waveguide can be used as the memory/amplification medium. The strong, broadband and single-mode light-atom interactions in such platforms are suitable for carrying out the amplification protocol proposed here.   
	
	 Alternatively, the loss processes can be mitigated   by using a cavity to modify the emission from the ensemble. For high-cooperativity light-atom interactions inside an optical resonator, the ratio of light emitted into the cavity compared to the free-space is enhanced by the cooperativity factor given by $ g^2/ ( \kappa \gamma) $ where $g$ is the light-atom coupling strength in the cavity, and $\kappa$ and $\gamma$ are the cavity and atomic decay rates.  By enhancing the interaction strength in an atom-cavity system, the efficiency of creating correlated Stokes and anti-Stokes photons has shown enhancement by about three orders of magnitude \cite{vuleticscience} compared to the free-space case \cite{Dou2018}. In macroscopic high finesse cavities with large optical access \cite{ PhysRevLett.118.183601,  Beck9740, Hosseini:PRL2016} simultaneous storage and projective measurement by single photon detection has been demonstrated for cooperativities higher than one. Using such ensemble cavity interactions, an experimental gain of $>3$ with 3\% success probability has already been observed \cite{Duan:Vuletic19}.  It was experimentally demonstrated that The state-of-the-art experiments with very high cooperaitvities and optical access \cite{PhysRevA.99.013437 , qinterface2012} are suitable platforms for realization of the proposed quantum amplification protocols. Overall, higher success probability and lower noise in the cavity regime is expected and further theoretical investigations are needed to quantify such dynamics to realized high quality amplifiers as outlined in Ref.\cite{caves2013}.

\section{\label{sec:level5} Summary}
	
	In summary, we have introduced an atomic state amplification operator, $S S^{\dagger}$, and proposed and examined the scheme to achieve probabilistic amplification. We have shown that by coherently adding excitations in an ensemble of atoms the probabilistic amplification of atomic coherence can be achieved through a process of Raman scattering forming $S S^{\dagger}$ operation. Also, we have shown that multiple amplification operations are feasible using $(S S^{\dagger})^n$  or $S^n {S^{\dagger}}^n$ process, with the former process being more efficient than the later, as long as $\alpha\ll1$.  As coherent mapping from/to optical to/from collective atomic states is well known, the amplification protocol can be applied to optical states after storing optical information as a stationary atomic coherence. The finding of this study may be used to enhance the performance of future quantum communication and sensing networks.  Especially when cavity interactions are considered, the quality of the amplifier is expected to improve. With further theoretical analysis as future work, we expect that experimental realization of the probabilistic amplification using an ensemble of atoms confined in the mode of a high optical Q cavity \cite{vuleticscience, Duan:Vuletic19}, or optical fibers \cite{Sayrin:15} with high success probability is within reach.  \\

\section*{Acknowledgments}
	The authors would like to acknowledge support of this work by the AFOSR grant number FA 9550-19-1-0371. We would like to thank Michael Hush for enlightening discussions.

\bibliography{refs}

\begin{thebibliography}{10}
\newcommand{\enquote}[1]{``#1''}

\bibitem{Ourjoumtsev:2007aa}
A.~Ourjoumtsev, H.~Jeong, R.~Tualle-Brouri, and P.~Grangier,
  \enquote{Generation of optical `schr{\"o}dinger cats'from photon number
  states,} {\protect\JournalTitle{Nature}} \textbf{448}, 784 EP -- (2007).

\bibitem{Bimbard:2010aa}
E.~Bimbard, N.~Jain, A.~MacRae, and A.~I. Lvovsky, \enquote{Quantum-optical
  state engineering up to the two-photon level,} {\protect\JournalTitle{Nature
  Photonics}} \textbf{4}, 243 EP -- (2010).

\bibitem{McConnell:2015aa}
R.~McConnell, H.~Zhang, J.~Hu, S.~{\'C}uk, and V.~Vuleti{\'c},
  \enquote{Entanglement with negative wigner function of almost 3,000 atoms
  heralded by one photon,} {\protect\JournalTitle{Nature}} \textbf{519}, 439 EP
  -- (2015).

\bibitem{Deleglise:2008aa}
S.~Del{\'e}glise, I.~Dotsenko, C.~Sayrin, J.~Bernu, M.~Brune, J.-M. Raimond,
  and S.~Haroche, \enquote{Reconstruction of non-classical cavity field states
  with snapshots of their decoherence,} {\protect\JournalTitle{Nature}}
  \textbf{455}, 510 EP -- (2008).

\bibitem{PhysRevLett.112.170501}
N.~Roch, M.~E. Schwartz, F.~Motzoi, C.~Macklin, R.~Vijay, A.~W. Eddins, A.~N.
  Korotkov, K.~B. Whaley, M.~Sarovar, and I.~Siddiqi, \enquote{Observation of
  measurement-induced entanglement and quantum trajectories of remote
  superconducting qubits,} {\protect\JournalTitle{Phys. Rev. Lett.}}
  \textbf{112}, 170501 (2014).

\bibitem{Pfaff:2012aa}
W.~Pfaff, T.~H. Taminiau, L.~Robledo, H.~Bernien, M.~Markham, D.~J. Twitchen,
  and R.~Hanson, \enquote{Demonstration of entanglement-by-measurement of
  solid-state qubits,} {\protect\JournalTitle{Nature Physics}} \textbf{9}, 29
  EP -- (2012).

\bibitem{Kimble4QM}
K.~S. Choi, A.~Goban, S.~B. Papp, S.~J. van Enk, and H.~J. Kimble,
  \enquote{Entanglement of spin waves among four quantum memories,}
  {\protect\JournalTitle{Nature}} p. doi:10.1038/nature09568 (2010).

\bibitem{ralund2008}
T.~C. Ralph and A.~P. Lund, \enquote{Nondeterministic noiseless linear
  amplification of quantum systems,} {\protect\JournalTitle{AIP Conference
  Proceedings}} p. 155 (2009).

\bibitem{pegg}
D.~T. Pegg, L.~S. Phillips, and S.~M. Barnett, \enquote{Optical state
  truncation by projection synthesis,} {\protect\JournalTitle{Phys. Rev.
  Lett.}} \textbf{81}, 1604 (1998).

\bibitem{xiang}
G.~Y. Xiang, T.~C. Ralph, A.~P. Lund, N.~Walk, and G.~J. Pryde,
  \enquote{Heralded noiseless linear amplification and distillation of
  entanglement,} {\protect\JournalTitle{Nature Photonics}} \textbf{4}, 316
  (2010).

\bibitem{zava}
A.~Zavatta, J.~Fiur\'{a}\v{s}ek, and M.~Bellini, \enquote{A high-fidelity
  noiseless amplifier for quantum light states,} {\protect\JournalTitle{Nature
  Photonics}} \textbf{5}, 52 (2011).

\bibitem{caves2013}
S.~Pandey, Z.~Jiang, J.~Combes, and C.~M. Caves, \enquote{Quantum limits on
  probabilistic amplifiers,} {\protect\JournalTitle{Phys. Rev. A}} \textbf{88},
  033852 (2013).

\bibitem{Barnett:09}
S.~M. Barnett and S.~Croke, \enquote{Quantum state discrimination,}
  {\protect\JournalTitle{Adv. Opt. Photon.}} \textbf{1}, 238--278 (2009).

\bibitem{PhysRevA.93.062315}
M.~Rosati, A.~Mari, and V.~Giovannetti, \enquote{Coherent-state discrimination
  via nonheralded probabilistic amplification,} {\protect\JournalTitle{Phys.
  Rev. A}} \textbf{93}, 062315 (2016).

\bibitem{Chrzanowski:2014aa}
H.~M. Chrzanowski, N.~Walk, S.~M. Assad, J.~Janousek, S.~Hosseini, T.~C. Ralph,
  T.~Symul, and P.~K. Lam, \enquote{Measurement-based noiseless linear
  amplification for quantum communication,} {\protect\JournalTitle{Nature
  Photonics}} \textbf{8}, 333 EP -- (2014).

\bibitem{dicke1954}
R.~H. Dicke, \enquote{Coherence in spontaneous radiation processes,}
  {\protect\JournalTitle{Phys. Rev.}} \textbf{93}, 99--110 (1954).

\bibitem{Tanji:PRL2009}
H.~Tanji, S.~Ghosh, J.~Simon, B.~Bloom, and V.~Vuleti\`c, \enquote{Heralded
  single-magnon quantum memory for photon polarization states,}
  {\protect\JournalTitle{Phys. Rev. Lett.}} \textbf{103}, 043601 (2009).

\bibitem{vuleticscience}
J.~K. Thompson, J.~Simon, H.~Loh, and V.~Vuleti\'{c}, \enquote{A
  high-brightness source of narrowband, identical-photon pairs,}
  {\protect\JournalTitle{Science}} \textbf{313}, 74 (2006).

\bibitem{dlcz}
L.~M. Duan, M.~D. Lukin, J.~I. Cirac, and P.~Zoller, \enquote{Long-distance
  quantum communication with atomic ensembles and linear optics,}
  {\protect\JournalTitle{NATURE}} \textbf{414}, 413 (2001).

\bibitem{Vernaz-Gris:2018aa}
P.~Vernaz-Gris, K.~Huang, M.~Cao, A.~S. Sheremet, and J.~Laurat,
  \enquote{Highly-efficient quantum memory for polarization qubits in a
  spatially-multiplexed cold atomic ensemble,} {\protect\JournalTitle{Nat.
  Commun.}} \textbf{9}, 363 (2018).

\bibitem{RamanQM}
K.~F. Reim, J.~Nunn, V.~O. Lorenz, B.~J. Sussman, K.~Lee, N.~K. Langford,
  D.~Jaksch, and I.~A. Walmsley, \enquote{Towards high-speed optical quantum
  memories,} {\protect\JournalTitle{arXiv:0912.2970v1, [quant-ph]}}  (2009).

\bibitem{Hosseini:NatPhys:2011}
M.~Hosseini, B.~M. Sparkes, G.~Campbell, P.~K. Lam, and B.~C. Buchler,
  \enquote{Unconditional room temperature quantum memory,}
  {\protect\JournalTitle{Nat. Phys.}} \textbf{7}, 794--798 (2011).

\bibitem{Gisin:Nature2007}
N.~Gisin and R.~Thew, \enquote{Quantum communication,}
  {\protect\JournalTitle{Nature Photonics}} \textbf{1}, 165 (2007).

\bibitem{Kimble:Nature2008}
H.~J. Kimble, \enquote{The quantum internet,} {\protect\JournalTitle{Nature}}
  \textbf{453}, 1023--1030 (2008).

\bibitem{Liao:2017aa}
S.-K. Liao \emph{et~al.}, \enquote{Satellite-to-ground quantum key
  distribution,} {\protect\JournalTitle{Nature}} \textbf{549}, 43--+ (2017).

\bibitem{Fleischhauer-EIT}
M.~Fleischhauer, A.~Imamoglu, and J.~P. Marangos, \enquote{Electromagnetically
  induced transparency: Optics in coherent media,} {\protect\JournalTitle{Rev.
  of Mod. Phys.}} \textbf{77}, 634 (2005).

\bibitem{Reim:2010p11902}
K.~F. Reim, J.~Nunn, V.~O. Lorenz, B.~J. Sussman, K.~C. Lee, N.~K. Langford,
  D.~Jaksch, and I.~A. Walmsley, \enquote{Towards high-speed optical quantum
  memories,} {\protect\JournalTitle{Nature Photonics}} \textbf{4}, 218 (2010).

\bibitem{Afzelius:2009p11906}
M.~Afzelius, C.~Simon, H.~de~Riedmatten, and N.~Gisin, \enquote{Multimode
  quantum memory based on atomic frequency combs,} {\protect\JournalTitle{Phys.
  Rev. A}} \textbf{79}, 052329 (2009).

\bibitem{Hedges:2010p11910}
M.~Hedges, J.~Longdell, Y.~Li, and M.~Sellars, \enquote{Efficient quantum
  memory for light,} {\protect\JournalTitle{Nature}} \textbf{465}, 1052--1056
  (2010).

\bibitem{Pu:2017aa}
Y.~F. Pu, N.~Jiang, W.~Chang, H.~X. Yang, C.~Li, and L.~M. Duan,
  \enquote{Experimental realization of a multiplexed quantum memory with 225
  individually accessible memory cells,} {\protect\JournalTitle{Nat. Commun.}}
  \textbf{8}, 15359 (2017).

\bibitem{Parniak:2017aa}
M.~Parniak, M.~Dabrowski, M.~Mazelanik, A.~Leszczynski, M.~Lipka, and
  W.~Wasilewski, \enquote{Wavevector multiplexed atomic quantum memory via
  spatially-resolved single-photon detection,} {\protect\JournalTitle{Nat.
  Commun.}} \textbf{8} (2017).

\bibitem{group}
A.~B. Klimov and S.~M. Chumakov, \emph{A Group Theoretical Approach to Quantum
  Optics} (Wiley-VCH Verlag GmbH $\&$ Co. KGaA, 2009), 1st ed.

\bibitem{loudon}
R.~Loudon, \emph{The Quantum Theory of Light} (Oxford Science Publishers,
  2000), 3rd ed.

\bibitem{eisaman-eit-sph}
M.~D. Eisaman, A.~Andre, F.~Massou, M.~Fleischhauer, A.~S. Zibrov, and M.~D.
  Lukin, \enquote{Electromagnetically induced transparency with tunable
  single-photon pulses,} {\protect\JournalTitle{Nature}} \textbf{438}, 837
  (2005).

\bibitem{3d}
L.~M. Duan, J.~I. Cirac, and P.~Zoller, \enquote{Three-dimensional theory for
  interaction between atomic ensembles and free-space light,}
  {\protect\JournalTitle{Phys. Rev. A}} \textbf{66}, 023818 (2002).

\bibitem{bohnet2012steady}
J.~G. Bohnet, Z.~Chen, J.~M. Weiner, D.~Meiser, M.~J. Holland, and J.~K.
  Thompson, \enquote{A steady-state superradiant laser with less than one
  intracavity photon,} {\protect\JournalTitle{Nature}} \textbf{484}, 78 (2012).

\bibitem{goban2015superradiance}
A.~Goban, C.-L. Hung, J.~Hood, S.-P. Yu, J.~Muniz, O.~Painter, and H.~Kimble,
  \enquote{Superradiance for atoms trapped along a photonic crystal waveguide,}
  {\protect\JournalTitle{Physical review letters}} \textbf{115}, 063601 (2015).

\bibitem{scheibner2007superradiance}
M.~Scheibner, T.~Schmidt, L.~Worschech, A.~Forchel, G.~Bacher, T.~Passow, and
  D.~Hommel, \enquote{Superradiance of quantum dots,}
  {\protect\JournalTitle{Nature Physics}} \textbf{3}, 106 (2007).

\bibitem{sorensen2016coherent}
H.~S{\o}rensen, J.-B. B{\'e}guin, K.~Kluge, I.~Iakoupov, A.~S{\o}rensen,
  J.~M{\"u}ller, E.~Polzik, and J.~Appel, \enquote{Coherent backscattering of
  light off one-dimensional atomic strings,} {\protect\JournalTitle{Physical
  review letters}} \textbf{117}, 133604 (2016).

\bibitem{Langbecker:2018aa}
M.~Langbecker, R.~Wirtz, F.~Knoch, M.~Noaman, T.~Speck, and P.~Windpassinger,
  \enquote{Highly controlled optical transport of cold atoms into a hollow-core
  fiber,} {\protect\JournalTitle{New J. Phys.}} \textbf{20} (2018).

\bibitem{Dou2018}
J.-P. Dou, A.-L. Yang, M.-Y. Du, D.~Lao, H.~Li, X.-L. Pang, J.~Gao, L.-F. Qiao,
  H.~Tang, and X.-M. Jin, \enquote{Direct observation of broadband nonclassical
  states in a room-temperature light--matter interface,}
  {\protect\JournalTitle{npj Quantum Information}} \textbf{4}, 31 (2018).

\bibitem{PhysRevLett.118.183601}
M.~Hosseini, Y.~Duan, K.~M. Beck, Y.-T. Chen, and
  V.~Vuleti\ifmmode~\acute{c}\else \'{c}\fi{}, \enquote{Cavity cooling of many
  atoms,} {\protect\JournalTitle{Phys. Rev. Lett.}} \textbf{118}, 183601
  (2017).

\bibitem{Beck9740}
K.~M. Beck, M.~Hosseini, Y.~Duan, and V.~Vuleti{\'c}, \enquote{Large
  conditional single-photon cross-phase modulation,}
  {\protect\JournalTitle{Proceedings of the National Academy of Sciences}}
  \textbf{113}, 9740--9744 (2016).

\bibitem{Hosseini:PRL2016}
M.~Hosseini, K.~M. Beck, Y.~Duan, W.~Chen, and V.~Vuleti\ifmmode~\acute{c}\else
  \'{c}\fi{}, \enquote{Partially nondestructive continuous detection of
  individual traveling optical photons,} {\protect\JournalTitle{Phys. Rev.
  Lett.}} \textbf{116}, 033602 (2016).

\bibitem{Duan:Vuletic19}
Y.~Duan, M.~Hosseini, K.~M. Beck, and V.~Vuleti{\'c}, \enquote{Heralded
  interaction control between quantum systems,}
  {\protect\JournalTitle{arXiv:1909.12826 [quant-ph]}}  (2019).

\bibitem{PhysRevA.99.013437}
A.~Kawasaki, B.~Braverman, E.~Pedrozo-Pe\~nafiel, C.~Shu, S.~Colombo, Z.~Li,
  O.~\"Ozel, W.~Chen, L.~Salvi, A.~Heinz, D.~Levonian, D.~Akamatsu, Y.~Xiao,
  and V.~Vuleti\ifmmode~\acute{c}\else \'{c}\fi{}, \enquote{Geometrically
  asymmetric optical cavity for strong atom-photon coupling,}
  {\protect\JournalTitle{Phys. Rev. A}} \textbf{99}, 013437 (2019).

\bibitem{qinterface2012}
A.~Stute, B.~Casabone, and B.~e.~a. Brandst{\"a}tter, \enquote{Toward an
  ion--photon quantum interface in an optical cavity,}
  {\protect\JournalTitle{Appl. Phys. B}} \textbf{107}, 13 (2012).

\bibitem{Sayrin:15}
C.~Sayrin, C.~Clausen, B.~Albrecht, P.~Schneeweiss, and A.~Rauschenbeutel,
  \enquote{Storage of fiber-guided light in a nanofiber-trapped ensemble of
  cold atoms,} {\protect\JournalTitle{Optica}} \textbf{2}, 353--356 (2015).

\end{thebibliography}









\end{document}